\documentstyle[12pt]{article}

\def\lromn#1{\uppercase\expandafter{\romannumeral#1}}

\begin{document}
\begin{flushright}
TU/96/510 \\
hep-th/9609223 \\
Revised.
\end{flushright}

\vspace{12pt}

\begin{center}
\begin{large}

\bf{
Quantum Dissipation and Decay in Medium
}
\end{large}

\vspace{36pt}
\begin{large}
I. Joichi, Sh. Matsumoto, and
M. Yoshimura

Department of Physics, Tohoku University\\
Sendai 980-77 Japan\\
\end{large}

\end{center}

\begin{center}
\vspace{54pt}

{\bf ABSTRACT}
\end{center}

Quantum dissipation in thermal environment is investigated,
using the path integral approach.
The reduced density matrix of the harmonic oscillator system 
coupled to thermal bath of oscillators is derived
for arbitrary spectrum of bath oscillators.
Time evolution and the end point of two-body decay of unstable particles
is then elucidated: After early transient times unstable particles
undergo the exponential decay, followed by the power law decay
and finally ending in a mixed state of residual particles 
containing contributions from both on and off the mass shell,
whose abundance does not suffer from the Boltzmann suppression.


\newpage

It is a long standing problem in physics to resolve how a quantum
system evolves with time under influence of an environment described by
some mixed state.
The effect of the environment on the system dynamics is governed by the
fluctuation (or the noise) and the dissipation, related by
the fluctuation-dissipation theorem.
The simplest, yet the most fundamental model of quantum dissipation
is harmonic oscillator coupled to infinitely many oscillators that
make up a bath in a mixed state. Two powerful methods to analyze
this problem are the quantum Langevin equation \cite{q-dissipation langevin}, 
and the path integral approach \cite{feynman-vernon}, 
\cite{q-dissipation path}, \cite{qbm review}.

The important and the difficult part of analysis of this problem is
how to treat the non-local part of the correlation of the system
variable obtained after integrating out environment variables. 
The non-locality is required from quantum mechanical principles.
Many past studies have relied on the local form of the dissipation kernel
in the path integral approach.
Some exceptions are consideration in limited applications such as
\cite{leggett 84}, \cite{ford-lewis-oconnell}.
It has been pointed out that this often-used
approximation of the local friction (in the path integral sense) has
a fundamental flaw; violation of the positivity of the density matrix
\cite{ambegaokar}.
This has made it difficult to extract in a reliable way
a fully quantum mechanical time
evolution of the small system immersed in thermal environment at early
times, especially when the temperature is low.
Needless to say, the approximation of the local friction is excellent
in the most important phase of time evolution, thus is useful in
many practical applications. But as we shall see, there are physical effects
that can be understood only with the full non-locality incorporated.

In the present work we extend previous works 
on the non-local dissipation \cite{qbm review}
and give a quantum mechanical formulation of the system evolution 
in thermal environment.
The present model itself is not new, and
indeed, is a standard model of system-environment interaction
discussed repeatedly in the quantum dissipation problem.
A new method is developed by fully exploiting analyticity familiar in
scattering theory, to treat the general class of environment models
characterized by
the spectral weight consistent with general principles, but otherwise
taken arbitrary.
We then apply this formalism 
to the two-body decay of unstable particle in field theory,
by identifying as the environment oscillator 
a composite operator of two decaying particles.
We may thus discuss details of the decay law and what is left behind 
after decay.
It is the extended analyticity of a kernel (denoted later by $F(z)$)
that allows us to treat the whole time region of the decay
process in a unified way.

Another difference of our study from most of the past ones is
our assumption of the presence of a gap in the spectrum 
of environment oscillators.
Although this appears a minor technical point, it actually leads to
differences at low temperatures and also for behaviors of
physical quantities at very late times. 
The gapless case often appears in condensed matter physics,
for instance phonons in medium. But there are cases in which one cannot
neglect the gap such as in the unstable particle decay if the mass of
the daughter particle is finite.

Let the system variable in question be denoted by $q$ and 
the environment variable by $Q_{a}$. 
The Lagrangian of our problem consists of three parts: 
\begin{equation}
L = L_{q}[q] + L_{{\rm int}}[q\,, Q] + L_{Q}[Q] \,.
\end{equation}
We take for the system-environment interaction the bilinear term with
infinitely many harmonic oscillators for the environment:
\begin{eqnarray}
&&
L_{{\rm int}} = -\, q\,\sum_{a}\,c_{a}Q_{a} \,,
\\ &&
L_{q} = \frac{1}{2}\, (\,\dot{q}^{2} - \omega _{0}^{2}\,q^{2}\,) \,, 
\\ &&
L_{Q} = \frac{1}{2}\, \sum_{a}\,(\,\dot{Q}_{a}^{2} - \omega _{a}^{2}\,
Q_{a}^{2}\,) \,.
\end{eqnarray}
The renormalization of the frequency will be discussed later in 
appropriate places.
Introduction of a counter term in relation to the renormalization
as often discussed in the literature is actually the problem of how
to relate the parameters of the theory such as $\omega _{0}$ to
observable quantities. We shall discuss the relevant observable parameter 
$\bar{\omega }$ later.

The response of the environment to the system is characterized by
what we call the response weight, denoted by $r(\omega )$:
For the dynamical system of harmonic oscillator
of frequency $\omega _{0}$ coupled to environment oscillators 
of discrete frequency $\omega _{a}$, it is
\begin{eqnarray}
r(\omega ) = \sum_{a}\,\frac{c_{a}^{2}}{2\omega _{a}}
\,\delta (\omega - \omega _{a})
 =
\int_{\omega _{c}}^{\infty }\,d\omega \,
\frac{c^{2}(\omega )}{2\omega }\,D(\omega )
\,,
\end{eqnarray}
where in the continuum limit $D(\omega )$ is the density of states 
per unit frequency.
$\omega _{c}$ is the threshold for $r(\omega ) \neq 0 $.
Although we often write expressions in the discrete distribution of
frequencies, we have to consider the continuum limit.
It is often useful to extend $r(\omega )$ to the region $\omega < 0$ by
\( \:
r(-\omega ) = - r(\omega ) \,.
\: \)

It is very useful to introduce the influence functional 
${\cal F}[\,q(\tau )\,, q'(\tau )\,]$, following, but
slightly modifying, the original definition of 
Feynman and Vernon \cite{feynman-vernon}. 
The basic idea is that one is interested in the
behavior of the $q-$system alone
and traces out the environment variable altogether.
This way one can compute the reduced density matrix for the system
that contains all informations without knowing too much details of
the environemnt.
We define the influence functional by convoluting with the initial
state of the environment unlike the original one. 
Once the influence functional is known, one may compute the transition
probability and any physical quantities of the $q-$system
by convoluting dynamics of the system under study. 
For instance, the transition probability is given, with introduction
of the density matrix $\rho ^{(R)}$, by
\begin{eqnarray}
&&
\int\,{\cal D}q(\tau )\,\int\,{\cal D}q'(\tau )\,
\int\,dq_{i}\,\int\,dq'_{i}\,\int\,dq_{f}\,\int\,dq'_{f} \nonumber \\
&& 
\psi ^{*}(q_{f})\,\psi ^{*}(q'_{i})\,{\cal F}[\,q(\tau )\,, q'(\tau )\,]\,
e^{iS[q] - iS[q']}\,
\psi (q_{i})\,\psi (q'_{f}) \nonumber 
\\ &&
\hspace*{0.5cm} 
\equiv 
\int\,dq_{f}\,\int\,dq'_{f}\,\psi ^{*}(q_{f})\,\psi (q'_{f})
\rho ^{(R)}(q_{f} \,, q'_{f}) \,, 
\end{eqnarray}
where $\psi $'s are wave functions of the initial and
the final $q-$states, and $S[q]$ is the action of the $q-$system.

The form of the influence functional is dictated by general principles
such as probability conservation and causality. Feynman and Vernon 
found a closed quadratic form consistent with these,
\begin{eqnarray}
{\cal F}[\,q(\tau )\,, q'(\tau )\,] &=& \nonumber \\
&& \hspace*{-3cm}
\exp \left[\,-\,\int_{0}^{t }\,d\tau \,\int_{0}^{\tau }\,ds\,
\left( \,\xi (\tau )\alpha_{R}(\tau - s)\xi (s) + i\,\xi (\tau )
\alpha _{I}(\tau - s)X(s)\,\right)\,\right] \,, 
\label{influence-f def} \\
&& \hspace*{-2cm}
{\rm with} \;
\xi (\tau ) = q(\tau ) - q'(\tau ) \,, \hspace{0.5cm} 
X(\tau ) = q(\tau ) + q'(\tau ) \,.
\end{eqnarray}
Thus two real functions $\alpha _{i}(\tau )$ are all we need to
characterize the system-environment interaction.
These are defined here in the range of 
\( \:
\tau \geq 0 \,.
\: \)
The fact that $\alpha_{i} $ depends on the difference of time
variables, $\tau - s$, is due to the assumed stationarity of the environment.

The correlation kernels appear in the influence functional
as a form of the nonlocal interaction and they
are the dissipation $\alpha _{I}$ and
the noise $\alpha _{R}$, related to the response weight by
\begin{eqnarray}
\alpha _{I}(\tau ) &=& -\,\frac{i}{2}\,\int_{-\infty }^{\infty }\,
d\omega \,r(\omega )\,e^{-\,i\omega \tau } \,, 
\\
\alpha _{R}(\tau ) &=& \frac{1}{2}\, \int_{-\infty }^{\infty }\,d\omega \,
\coth (\frac{\beta \omega }{2})\,r(\omega )\,e^{-\,i\omega \tau } \,.
\end{eqnarray}

In the path integral approach integration over the sum variable
$X(\tau )$ is trivial for the harmonic oscillator system, 
since both the local part and the nonlocal action above
are linear in this variable:
\begin{eqnarray}
&&
\frac{1}{2}\, \int_{0}^{t}\,\left( \,\dot{\xi }(\tau )\dot{X}(\tau )
- \omega _{0}^{2}\,\xi (\tau )X(\tau )\,\right) - \nonumber 
\\ && \hspace*{-1cm}
\int_{0}^{t }\,d\tau \,\int_{0}^{\tau }\,ds\,
\left( \,\xi (\tau )\alpha_{R}(\tau - s)\xi (s) + i\,\xi (\tau )
\alpha _{I}(\tau - s)X(s) \,\right) \,. \nonumber 
\end{eqnarray}
Thus result of the path integration
of the system variable $X(\tau )$ gives the classical 
integro-differential equation that $\xi (\tau )$ must obey
\begin{equation}
\frac{d^{2}\xi }{d\tau ^{2}} + \omega^{2} _{0}\,\xi (\tau ) +
2\,\int_{\tau }^{t}\,ds \,\xi (s)\,\alpha _{I}(s - \tau) = 0 \,.
\label{xi integro-eq} 
\end{equation}
The end result of the $\xi $ path integral then 
contains an integral of the form,
\begin{eqnarray}
-\,
\int_{0}^{t}\,d\tau \,\int_{0}^{\tau }\,ds\,\xi (\tau )\alpha _{R}(\tau - s)
\xi (s) \,,
\end{eqnarray}
using the classical solution $\xi (\tau )$ with specified boundary
conditions,
\( \:
\xi (0) = \xi _{i} \,, \hspace{0.3cm} \xi (t) = \xi _{f}
\: \).

In the local approximation often used the dissipation kernel of the form
is taken to have the form of
\begin{eqnarray}
\alpha _{I}(\tau ) = \delta \omega ^{2}\,\delta (\tau ) 
+ \eta \,\delta '(\tau ) \,, 
\end{eqnarray}
with $\delta \omega ^{2}$ the frequency shift and the $\eta $ term
representing the local friction.
This choice enables one to solve the $\xi $ equation (\ref{xi integro-eq}) 
by elementary means.
On the other hand, the noise kernel is given by the response weight of
the form,
\begin{equation}
r(\omega ) = \frac{\eta }{\pi }\,\omega f(\frac{\omega }{\Omega }) \,, 
\end{equation}
with $f(x)$ some cutoff function and $\Omega $ a high frequency cutoff.
At high temperatures this approximation 
reduces to the well known classical form of the fluctuation,
\begin{equation}
\alpha _{R}(\tau ) = \frac{\eta T}{\pi }\,\delta (\tau ) \,.
\end{equation}
At low temperatures, howerver,
the use of the cutoff function only in the noise
kernel while retaining the local form of the dissipation
makes validity of this approximation dubious.
Our present approach does not make this local approximation, and
instead uses exact solutions of the classical equation (\ref{xi integro-eq}).
This equation has been used in other approaches 
\cite{leggett 84}, \cite{ford-lewis-oconnell}, \cite{qbm review}, too.

Two observations are crucial for our subsequent development: 
First, the integro-differential equation (\ref{xi integro-eq}) 
is converted to the one having the Volterra type
kernel with the variable change 
\( \:
\tau \rightarrow t - \tau \,. 
\: \)
It can then be solved by the Laplace transform, if the Laplace transform
of $\alpha _{I}(\tau )$ denoted by $\tilde{\alpha }(p)$ has a simple form.
Second, in thermal environment 
\( \:
\alpha (\tau ) = \alpha _{R}(\tau ) + i\alpha _{I}(\tau )
\: \)
is identified \cite{hjmy 96} to the real-time thermal Green's function, 
hence its Fourier transform has the well known analyticity property.
The origin of the analyticity is ultimately traced to the causality:
any physical disturbance cannot propagate faster than the light.

Let us explain this in more detail.
First, the Laplace transformed kernel of evolution operator is
\begin{eqnarray}
\tilde{g}(p) = \frac{1}{p^{2} + \omega _{0}^{2} + 2\tilde{\alpha }(p)} \,,
\end{eqnarray}
with real $p>0$.
\begin{eqnarray}
\tilde{\alpha }(p) = 
\frac{1}{2}\, \int_{-\infty }^{\infty }\,d\omega \,
\frac{r(\omega )}{ip - \omega }
\end{eqnarray}
is equal to the boundary value of the analytic function 
$\pi \overline{G}(ip)$.
This analytic function $\overline{G}(z)$ is related \cite{hjmy 96} 
to the imaginary-time thermal Green's function \cite{fetter-walecka} 
and is regular in the complex
$z$ plane except on the real axis where it has the branch point singularity
with the discontinuity given by $r(\omega )$.
Our notational convention here is that $\omega $ and $p$ are real and
the complex variable is denoted by $z$.

We next introduce the analytically extended function by
\begin{eqnarray}
&&
F(z) = \tilde{g}(-iz) = \frac{1}
{-z^{2} + \omega _{0}^{2} + 2\pi \overline{G}(z)} \,, 
\\ &&
\overline{G}(z) = \int_{- \infty }^{\infty }\,\frac{d\omega }{2\pi }\,
\frac{r(\omega )}{z - \omega } \,.
\end{eqnarray}
It can be proved from the positivity of the response weight $r(\omega )$ 
supplemented by the spectrum condition later explained that
$F^{-1}(z)$ has no complex zero.\cite{schwinger unstable} 
Hence $F(z)$ is regular except on the real axis.
Its discontinuity across the real axis
\begin{eqnarray}
H(\omega ) = \left( \,F(\omega + i\epsilon ) - F(\omega - i\epsilon )\,\right)
/(2\pi i)
\end{eqnarray}
is fundamental to our subsequent analysis.
In terms of two real functions,
\( \:
\Pi (\omega ) 
\: \)
and \( \:
\Gamma (\omega  )
\: \)
defined by
\begin{eqnarray}
&&
2\pi \,
\overline{G}(\omega + i\epsilon ) = \Pi (\omega ) - i\,\omega \Gamma (\omega )
\,, 
\\ &&
H(\omega ) 
=
\frac{1}{\pi }\,
\frac{\omega \Gamma (\omega )}{(\,\omega ^{2} - \omega _{0}^{2}
- \Pi (\omega )\,)^{2} + \omega^{2}\,\Gamma ^{2}(\omega )}
\,. 
\end{eqnarray}
Furthermore, 
\( \:
r(\omega ) = \omega \Gamma (\omega )/\pi   \,,
\: \)
and the real part $\Pi (\omega )$ may be written by the well known
dispersion integral:
\begin{eqnarray}
\Pi (\omega ) = {\cal P}\,\int_{-\infty }^{\infty }\,d\omega '\,
\frac{r(\omega ')}{\omega - \omega '}
\,,
\end{eqnarray}
with ${\cal P}$ denoting the principal value of integration.

As in scattering theory \cite{gunson-taylor}, the analytic function
$F(z)$ may be extended to the second
Riemann sheet through the discontinuity formula,
\begin{eqnarray}
\! 
F(\omega + i\epsilon ) - F(\omega - i\epsilon ) = i
2\pi r(\omega )F(\omega + i\epsilon )F(\omega - i\epsilon )
\,. \! 
\end{eqnarray}
In the second lower sheet one finds a finite number of poles at $z$ obeying
\begin{equation}
z ^{2} - \omega _{0}^{2} - 2\pi \,\overline{G}(z) 
+ 2\pi ir(z) = 0 \,.
\end{equation}
Except at these poles $F(z)$ is regular in the unphysical sheet.

Physical interpretation of these singularities is possible in terms of
the spectrum of the entire harmonic system.
Due to the mixing with the environment the system spectrum
is modified and is given by an eigenvalue equation for 
$\lambda = \omega ^{2}$.
When the response weight $r(\omega ) \rightarrow $ a constant 
at $\omega \rightarrow \infty $, as happens often and in field theory 
models later discussed in particular,
one has to subtract a term from integrals  containing the response weight.
This corresponds to renormalization of the bare frequency $\omega _{0}$ with
the frequency shift,
\begin{equation}
\delta \omega ^{2} = 
- \,2\,\int_{\omega _{c}}^{\infty }\,d\omega \,r(\omega )/\omega 
\,.
\end{equation}
The renormalized frequency is given by
\( \:
\omega _{R}^{2} = \omega _{0}^{2} + \delta \omega ^{2} \,.
\: \)
The eigenvalue equation written using $\omega _{R}^{2}$ is then
\begin{eqnarray}
\lambda - \omega _{R}^{2} - \lambda \,
\int_{\omega _{c}}^{\infty }\,d\omega \,\frac{2r(\omega )}
{\omega\, (\,\lambda - \omega ^{2}\,)} = 0
\,. \label{spectrum eigenvalue} 
\end{eqnarray}
The left hand side coincides with $-\,F^{-1}(\sqrt{\lambda })$.
The condition of stability of the entire system requires that 
the smallest eigenvalue $\lambda > 0$, giving
\( \:
\omega _{R}^{2} > 0
\,,
\: \)
or equivalently
\begin{equation}
\omega _{0}^{2} > 2\,\int_{\omega _{c}}^{\infty }\,d\omega \,r(\omega )/
\omega \,, 
\end{equation}
if the integral on the right hand side is convergent.
This is the condition needed to ensure the analyticity of $F(z)$
as described above.

We do not consider the possibility of an isolated pole of $F(z)$ 
on the real axis outside the cut, 
which corresponds to a stable
state, implying that in the problem raised here a prepared initial
configuration of the system does not completely decay.
We thus focus on physical systems that have a continuous spectrum alone
when interaction between the small system and the large environment
is considered.

The complex pole in the second Riemann sheet describes the behavior of
the system
oscillator interacting with environment oscillators: The real part
of the pole position $\bar{\omega }^{2}$
gives a physical frequency squared including
the frequency shift, and its imaginary part is related to the decay rate
of the system oscillator.
The pole {\em mass} here $\bar{\omega } \neq \omega _{R}$. 
Note that the pole position $\bar{\omega }$ is, but $\omega _{R}$
is not, an observable quantity.
We assume as a physical requirement that a single pole exists 
in the nearby second sheet, but it should be evident that
an arbitrary number of poles can readily be incorporated.

Subsequent computation frequently uses the Laplace inverted evolution operator
$g(\tau )$ which can be written in terms of $H(\omega )$:
\begin{eqnarray}
g(\tau ) =
2\,\int_{\omega_{c}}^{\infty }\,d\omega \,H(\omega )\sin (\omega \tau )
\,. 
\end{eqnarray}
Solution of the integro-differential equation (\ref{xi integro-eq}) 
is then given by
\begin{eqnarray}
\xi (\tau ) =
\xi _{i}\,\frac{g(t - \tau )}{g(t)} + \xi _{f}\,
\left( \,\dot{g}(t - \tau ) - \frac{g(t - \tau )\dot{g}(t)}{g(t)} \,\right)
\,, \nonumber 
\end{eqnarray}
with the dot denoting derivative.
Both $g(t)$ and $\dot{g}(t)$ can be shown to satisfy the integro-differential
equation of the form ($x = g \;$ or $\dot{g}$),
\begin{eqnarray}
\frac{d^{2}x}{dt^{2}} + \omega _{0}^{2}\,x + 2\,\int_{0}^{t}\,d\tau \,
\alpha _{I}(t - \tau  )\,x(\tau ) = 0 \,.
\end{eqnarray}

The reduced density matrix of the quantum system at any time is obtained from
the action written in terms of the boundary values,
\( \:
S_{cl}(\xi _{f} \,, X_{f} \, ; \, \xi _{i} \,, X_{i}) \,, 
\: \)
by convoluting with the initial density matrix.
This action is computed as
\begin{eqnarray}
i\,S_{{\rm cl}} &=& -\,\frac{U}{2}\,\xi _{f}^{2} - \frac{V}{2}\,\xi _{i}^{2}
- W\,\xi _{i}\,\xi _{f}  + 
\frac{i}{2}\,X_{f}\,\dot{\xi }_{f} - 
\frac{i}{2}\,X_{i}\,\dot{\xi }_{i} \,, 
\\
U &=& 2\,\int_{0}^{t }\,d\tau \,\int_{0}^{\tau }\,ds\,
z(\tau )\,\alpha _{R}(\tau - s)\,z(s) \,, 
\\
V &=& 2\,\int_{0}^{t }\,d\tau \,\int_{0}^{\tau }\,ds\,
y(\tau )\,\alpha _{R}(\tau - s)\,y(s) \,, 
\\
W &=& \int_{0}^{t }\,d\tau \,\int_{0}^{\tau }\,ds\,
\left( \,y(\tau )z(s) + y(s)z(\tau )\,\right)\,\alpha _{R}(\tau - s) \,, 
\\
y(\tau ) &=&
\frac{g(t - \tau )}{g(t)} \,, 
\\
z(\tau ) &=& \dot{g}(t - \tau ) - g(t - \tau )\frac{\dot{g}(t)}{g(t)} \,, 
\\
\dot{\xi }(\tau ) &=&
-\,\xi _{i}\frac{\dot{g}(t - \tau )}{g(t)} - \xi _{f}\,
\left( \,\stackrel{..}{g}(t - \tau ) - \frac{\dot{g}(t - \tau )\dot{g}(t)}
{g(t)}\,\right) \,.
\end{eqnarray}
The same effective action as ours has been derived by Grabert et al.,
as summarized in \cite{qbm review}. Our derivation here simplifies their
calculation, with a new form of the reduced density matrix, 
Eq.(\ref{density matrix}) that follows, 
by fully exploiting the extended analyticity.
Our method also makes possible a unified
treatment of the exponential and the power law decay, as will be made
clear later.

We take as the initial state a product of thermal states, a system of
temperature $T_{0} = 1/\beta _{0}$ 
and an environment of temperature $T = 1/\beta $.
We may take $T_{0} = T$ when we apply to the decay process of
excited level initially in thermal equilibrium.
In the limit of $T_{0} \rightarrow 0$ it describes the ground state
of the system harmonic oscillator.
(We have recently computed the reduced density matrix, starting from
another pure state, the first excited level of the system ocsillator.
As expected, the late time behavior in this case is the same as in the
present case.)

After a series of straightforward
Gaussian integration we find the reduced density matrix of the form,
\begin{eqnarray}
&&
\rho ^{(R)}(X_{f}\,, \xi _{f}) = 2\sqrt{\frac{{\cal A}}{\pi }}
\exp [\,-{\cal A}X_{f}^{2} - {\cal B}\xi _{f}^{2} +
i{\cal C}X_{f}\xi _{f}\,] \,,
\label{density matrix} 
\\
&&
{\cal A} = \frac{1}{8 I_{1}}\,, \hspace{0.5cm} 
{\cal B} = \frac{1}{2}\, (\,I_{3} - \frac{I_{2}^{2}}{I_{1}}\,) \,, 
\hspace{0.5cm} 
{\cal C} = \frac{I_{2}}{2I_{1}} \,, 
\\
&&
I_{1} =
{\cal I}[\,|h(\omega \,, t)|^{2}\,] + 
\frac{1}{2\omega _{0}}\coth (\frac{\beta _{0}\omega _{0}}{2})
(\dot{g}^{2} + \omega _{0}^{2}g^{2})
\,, \nonumber 
\\
&&
I_{2} =
\Re {\cal I}[\,h(\omega \,, t)k^{*}(\omega \,, t)\,] +
\frac{1}{2\omega _{0}}\coth (\frac{\beta _{0}\omega _{0}}{2})
\dot{g}(\stackrel{..}{g} + \omega _{0}^{2}g)
\,, \nonumber 
\\
&&
I_{3} =
{\cal I}[\,|k(\omega \,, t)|^{2}\,]
+ \frac{1}{2\omega _{0}}\coth (\frac{\beta _{0}\omega _{0}}{2})
(\stackrel{..}{g}^{2} + \omega _{0}^{2}\dot{g}^{2}) \,.
\end{eqnarray}
$\omega _{0}$ is a reference frequency taken as the initial system
state, and equated here to the initial system frequency. 
If one so desires, either the renormalized 
$\omega _{R}$ or the pole $\bar{\omega }$ may be taken as another
choice. But we imagine the situation a small system was added to a large
environment at some time, 
its mutual interaction being absent prior to the initial time.
In this circumstance it is appropriate to take $\omega _{0}$ as the reference
frequency.
As will be made clear later, dependence on the initial state dies away
quickly as time passes.
In writing the reduced density matrix, we introduced the following notations;
\begin{equation}
{\cal I}[\,f(\omega )\,] \equiv 
\int_{\omega _{c}}^{\infty }d\omega \,
\coth (\frac{\beta \omega }{2})r(\omega )f(\omega ) \,, 
\end{equation}
and 
\begin{eqnarray}
&&
h(\omega \,, t) \equiv \int_{0}^{t}d\tau \,g(\tau )e^{-\,i\omega \tau }
\,, 
\\ &&
k (\omega \,, t) 
\equiv \int_{0}^{t}d\tau \,\dot{g}(\tau )e^{-\,i\omega \tau } =
g(t)e^{-\,i\omega t} + i\omega h(\omega \,, t) \,.
\end{eqnarray}
The density matrix $\rho ^{(R)}$ from which any physical quantity
can be computed at any time has explicitly been given by the discontinuity,
$H(\omega )$ or $r(\omega )$.

This density matrix $\rho ^{(R)}$ is positive definite and behaves acceptably
at early times unlike the one in the local approximation, as will
be explained elsewhere \cite{lindblad}.
Our interest here is limited to the late time behavior.

It is important to know the behavior of the kernel $g(\tau )$,
which is found by deforming the contour of $\omega $ integration into
the sum of the pole contribution (at $z = z_{0} $  with $ \Im z_{0} < 0$)
in the second sheet and the contribution
parallel to the imaginary axis passing through $z = \omega _{c} $, both
in the first (\lromn 1) and in the second (\lromn 2)
sheet \cite{goldberger-watson}:
\begin{eqnarray}
&&
g(\tau ) =
\Im \left( Ke^{-i\Re z_{0}\tau } \right)\,
e^{\Im z_{0}\tau } \nonumber 
\\
&+& 
\Im \left[ \frac{e^{i\omega _{c}\tau }}{\pi }\int_{0}^{\infty }dy\,
e^{- y\tau }\left( \,F_{{\rm \lromn 1}}(\omega _{c} + iy) - F_{{\rm \lromn 2}}
(\omega _{c} + iy)\,\right) \right] \,, 
\end{eqnarray}
with
\( \:
K^{-1} = z_{0} - \pi \overline{G}'(z_{0} ) + i\pi r'(z_{0})\,.
\: \)

As seen from this formula, 
the pole contribution given by the first term describes
the exponential decay, which usually lasts 
during the most dominant phase of the decay period, 
while the rest of contribution gives the power law decay at very late times;
$\propto t^{-\,\alpha -1}$, and well-behaved early time behavior.
The power $-\,\alpha -1$ is dictated by the threshold behavior 
of the response weight,
\( \:
r(\omega ) \:\propto  \: (\omega - \omega _{c})^{\alpha } \,.
\: \)
$\alpha  = 1/2$ for the S-wave two-body decay of unstable particle
\cite{goldberger-watson}.
$\alpha  > - 1$ is required as a consistency of this approach;
convergence of the $\omega $ integration.

It is often claimed that the power law behavior is never observable, since
by the time this term dominates the exponential decay essentially
eliminates the initial population.
This is perhaps so when the environment is at zero temperature.
In order to check observability of the power law decay in medium,
we shall estimate the transition time $t_{*}$ from the
exponential to the power law period at low, but finite temperatures.
Let us examine a typical case by taking the form of the response
weight,
\( \:
r(\omega ) = c\,(\omega - \omega _{c})^{\alpha } \,, 
\: \)
with $0 < \alpha < 1$ in the range of $\omega _{c} < \omega < \Omega $
($\Omega \gg \omega _{c}$) and with
\( \:
\bar{\omega } \gg  {\rm Max}\;(\,\omega_{c} \,, T\,)
\: \).
The behavior of $g(t)$ in the power law period is
\begin{eqnarray}
g(t) \approx \frac{2c}{\bar{\omega }^{4}}\,\Gamma (\alpha + 1)\,
\frac{\cos (\,\omega _{c}\,t + \frac{\pi }{2}\,\alpha \,)}
{t^{\alpha + 1}} \,. \label{late time kernel} 
\end{eqnarray}
By equating this to the expression for the same quantity
$g(t)$ in the exponential period, one obtains
\begin{equation}
t_{*} \approx \frac{1}{\gamma }\,\ln \left( \frac{\bar{\omega }^{3}}
{2c\,\Gamma(\alpha + 1)\,\gamma ^{\alpha + 1}}\right)
 \,,
\end{equation}
with 
\( \:
\gamma = -\,\Im z_{0} \,
\: \)
the decay rate.
In the weak coupling limit the factor inside the logarithm is large
($\propto c^{-\,2\alpha - 3}$), and by the time $t_{*}$ the initial
population has decreased like
\begin{equation}
e^{-\,2\gamma t_{*}} \:\propto  \: c^{4\alpha + 6} \,.
\end{equation}
At non-zero temperatures the response weight $r(\omega )$ may depend
on the environment temperature, as will be made more explicit in
our application to the unstable particle decay, for instance in 
eq.(\ref{response for decay}). 
Temperature dependence of the parameters in these formulas is thus
needed to check observability of the power law decay at finite
temperatures.

Emergence of the power law term in the quantum Brownian motion
has been noted in some specific models of Ohmic type,
\cite{grabert-weiss}, \cite{haake et al}, \cite{power comment}, 
but we find this behavior as a general property 
in the presence of the non-local dissipation.
For instance, the power law behavior of the two point correlation
function was noted in ref. \cite{grabert-weiss}, \cite{haake et al}, 
for specific models.
We have instead demonstrated the power law behavior for the kernel
function $g(t)$ by separating the non-pole contribution that is essential
to the power law behavior.
Hence we have shown this property in more generality.
We would like to stress that the presence of the branch cut singularity
of the function $F^{-1}(z)$ in the
$z$ plane is very important to derive the power law decay.
This cut is further related to the non-analytic property of the
real function $r(\omega )$ at the threshold, including the special case
of the gapless  $\omega _{c} = 0$.
For instance, if one takes the Ohmic form of the response weight given
by pole terms alone, such as
\begin{eqnarray}
r ^{(4)}(\omega ) = \frac{4c\Omega \gamma\, \omega }
{(\omega^{2} - \Omega ^{2}
+ \frac{\gamma ^{2}}{4})^{2} + \Omega ^{2}\gamma ^{2}} \,, 
\end{eqnarray}
then one does not obtain the power law decay for $g(t)$.
For the gapless case of a fractional power $\alpha $ the formula
(\ref{late time kernel}) is still valid, but the kernel 
$g(t)$ does not exhibit
the power law behavior for an odd integer $\alpha $, as seen from
(\ref{late time kernel}) with $\omega _{c} = 0$.
In this case one can define a regular function $r(\omega )$ including
$\omega = 0$, thus the gapless case of odd integer $\alpha $
is exceptional in the sense that only this case does not produce
the power law behavior.
The critical condition for the absence of the power law behavior of
$g(t)$ is then regularity of $r(\omega )$ at the threshold.

The asymptotic late time behavior of the reduced density matrix
is determined by 
\( \:
h(\omega \,, \infty ) = \int_{0}^{\infty }\,
d\tau \,g(\tau )e^{-\,i\omega \tau } \,, 
\: \)
which is shown to be equal to the boundary value,
\( \:
F(\omega - i\epsilon ) \,,
\: \)
thus giving 
\begin{eqnarray}
&&
r(\omega )|h(\omega \,, \infty )|^{2} = H(\omega ) \,, 
\\ &&
r(\omega )|k(\omega \,, \infty )|^{2} = \omega ^{2}H(\omega )
\,.
\end{eqnarray}
In these computations the analytic structure and only that is
important. Thus, at asymptotic late times
\begin{eqnarray}
{\cal A} &\rightarrow  &
\frac{1}{8}\,
\left( \,
\int_{\omega _{c}}^{\infty }
d\omega \,\coth (\frac{\beta \omega }{2})\,H(\omega )\,\right)^{-1}
\,, \label{asymptotic a} 
\\
{\cal B} &\rightarrow &
\frac{1}{2}\,
\int_{\omega _{c}}^{\infty }d\omega \,\coth (\frac{\beta \omega }{2})\,
\omega ^{2}H(\omega )
\,, \label{asymptotic b} 
\\ 
{\cal C} &\rightarrow&    0 \,.
\end{eqnarray}
Note that the dependence on the initial state of the system via the
$\beta _{0} = 1/T_{0}$ factor disappears, 
hence no memory effect of the initial state remains in the asymptotic 
final state.

In the high temperature limit 
\begin{equation}
{\cal A} \sim \frac{\omega_{R}^{2}}{8T} \,, \hspace{0.5cm} 
{\cal B} \sim \frac{T}{2} \,.
\end{equation}
This result is almost equal to the pole contribution, igoring a minor
difference between $\omega _{R}$ and $\bar{\omega }$.
Contribution during the period of
the power law decay is given by the continuous integral and is numerically
subdominant, suppressed by the factor
$1/T^{2}$ relative to the one from the exponential period.

On the other hand, at low temperatures the contribution from
the threshold region, $\omega \approx \omega _{c}$, cannot be ignored,
giving the dominant contribution to the power law period.

Let us apply these considerations to the decay of unstable particle;
\( \:
\varphi \rightarrow \chi + \chi \,.
\: \)
We assume that the decay product $\chi $ is a part of thermal components that
make up the environment.
The parent particle $\varphi $ may or may not be in thermal equilibrium
with the rest of medium. Since we focus on the late time behavior,
the initial state dependence disappears.
Unlike the unstable particle decay in vacuum some amount of parent
particles are expected to be left behind, even much later than the decay
lifetime, because in thermal environment even heavier parent particles
can be created by energetic particles of smaller mass.
If so, what is the fraction of the parent particle left behind?

The interaction of the system field $\varphi $ and the environment
field $\chi $ is assumed to be described by a relativistic field theory 
of Yukawa interaction Lagrangian density,
\( \:
\frac{1}{2}\, \mu \varphi \chi ^{2} \,,
\: \)
with $\mu $ a coupling constant of mass dimension.
We presume that interaction among created $\chi $ particles themselves
is weak enough. In this case it is possible to identify as 
the environment variable the two-body bilinear operator of $\chi $,
\begin{equation}
\sum_{a}\,c_{a}Q_{a} = 
\frac{\mu }{2}\,\int\,d^{3}x\,\chi ^{2}(x)\,e^{-\,i\vec{k}\cdot \vec{x}}
\,, 
\end{equation}
with $\vec{k}$ the momentum of $\varphi $ particle.
Due to the assumed homogeneity each $\vec{k}$ mode can be treated
independently.
The continuous label $\omega $ in the environment variable $Q(\omega )$
is identified here to 
the internal configuration of two body states with a given total momentum
$\vec{k}$.
This field theory model was introduced in ref \cite{hjmy 96}.

The relevant response weight for $\varphi \rightarrow \chi + \chi $ has 
been given in ref \cite{hjmy 96}.
We shall recapitulate the main point of that calculation.
The important point is that to lowest non-trivial 
order of the coupling $\mu $,
$r(\omega )$ is given by the imaginary part of the one-loop diagram
of the self-energy of $\varphi $ at finite temperatures, $\varphi $ being
off the mass shell: 
\( \:
\omega ^{2} - \vec{k}^{2} \neq 
\: \)
the $\varphi $ mass$^{2}$.
The result for the $\chi $ loop diagram at finite temperatures is well known 
\cite{weldon}, \cite{hjmy 96},
and physically consists of two parts; the process $\varphi \leftrightarrow 
\chi + \chi $ in the region of
$\omega > \sqrt{k^{2} + 4m^{2}}$ and the other process
\( \:
\varphi + \chi \leftrightarrow \chi 
\: \)
(forbidden when all particles are on the mass shell, but allowed
in thermal environment)
in $0 < \omega < k$, where $m$ is the daughter $\chi $ mass.
The response weight $r(\omega )$ does not vanish for
\( \:
|\omega | > \sqrt{\vec{k}^{2} + 4m^{2}}
\: \)
and
\( \:
|\omega | < |\vec{k}|
\: \)
from the kinematics of the decay and the inverse decay of particles 
off the mass shell, with the constraint of the momentum conservation.
Thus a gap of the spectrum exists in
\( \:
k < |\omega | < \sqrt{\vec{k}^{2} + 4 m^{2}}
\,. 
\: \)
The finite non-vanishing mass of the daughter particle ($m \neq 0$)
is important for the existence of the gap and for associated
physical consequences that follow.

For $\omega > \sqrt{k^{2} + 4m^{2}}$ the response weight is
\begin{eqnarray}
r(\omega ) = \frac{\mu ^{2}}{32\pi^{2} }\,
\left( \,\sqrt{\,1 - \frac{4m^{2}}{\omega ^{2} - k^{2}}\,} +
\frac{2}{k\beta }\,\ln \frac{1 - e^{-\beta \omega _{+}}}
{1 - e^{-\beta |\omega _{-}|}}\,\right) \,, 
\nonumber \\ \label{response for decay}
\end{eqnarray}
with 
\begin{equation}
\omega _{\pm } = \frac{\omega }{2} \pm \frac{k}{2}\,
\sqrt{\,1 - 4m^{2}/(\,\omega ^{2} - k^{2}\,)\,} \,.
\end{equation}
For $0 < \omega < k$ only the second term in the bracket of 
Eq.(\ref{response for decay}) contributes.

A useful, and adequate approximation we exploit for subsequent estimate
is the weak coupling scheme with correct
threshold and asymptotic behaviors incorporated: 
\begin{equation}
F(z) =
\frac{1}{- z^{2} + \bar{\omega }^{2} - i\,\pi r(z)} \,.
\end{equation}
In this approximation we replaced the real part $\Pi (\omega )$ by
the constant pole location $\bar{\omega }$.

A quantity of physical interest is the fraction of remaining particles 
given by the occupation number at asymptotic late times, 
\begin{equation}
n_{k} = \frac{1}{2}\, \langle \,\frac{p_{k}^{2}}{\omega _{k}} +
\omega _{k}q_{k}^{2}\, \rangle \approx  \frac{{\cal B}_{k}}{\omega _{k}} + 
\frac{\omega _{k}}{16{\cal A}_{k}} - \frac{1}{2} \,, 
\end{equation}
for each $\vec{k}$ mode ($\omega _{k} = \sqrt{\vec{k}^{2} + M^{2}}$).
The temperature dependent part of this quantity is
\begin{equation}
n^{\beta  }_{k} = 
\int_{0}^{\infty }\,d\omega \,\frac{1}{e^{\beta \omega }
- 1}\,(\omega_{k}  + 
\frac{\omega ^{2}}{\omega_{k}} )\,H(\omega \,, k) \,, 
\end{equation}
with $\omega _{k}$ the real part of the pole position.
One must sum over $\vec{k}$ to obtain the number density of remnants.

We shall limit our discussion here to the decay that occurs 
when the parent $\varphi $ becomes non-relativistic, 
\begin{equation}
\omega_{k} \sim  M + \frac{\vec{k}^{2}}{2M} \gg T \,.
\end{equation}
This condition is relevant in interesting cosmological problems of the neutron
decay at the time of nucleosynthesis and GUT $X$ boson decay at
baryogenesis \cite{baryogenesis review96}.

Computation of the temperature dependent part of the occupation number
$n_{k}^{\beta }$ may proceed 
by deforming the contour of $\omega $ integration, in the same way
as in the discussion of $g(\tau )$ above. 
There are then two types of contribution: One is the pole term that gives 
the usual Boltzmann suppressed contibution of $e^{-\,\beta \omega _{k}}$.
When mode-summed, it gives the number density,
\begin{equation}
\int\,\frac{d^{3}k}{(2\pi )^{3}}\,e^{-\,\beta (\,M + k^{2}/2M\,)}
= (\frac{MT}{2\pi })^{3/2}\,e^{-M/T} \,.
\end{equation}
This is the familiar Boltzmann suppressed formula.

The second one is contribution from the continuous
complex path that gives the power law behavior of temperature dependence.
A part of this contribution in the region $\omega > \sqrt{k^{2} + 4m^{2}}$ 
is analytically calculable by using
\( \:
r(\omega ) = \frac{\mu ^{2}}{16\pi } + O[m^{2}] \,, 
\: \)
valid for a small daughter mass $m$. It is
\begin{eqnarray}
n &\approx& \frac{\mu ^{2}}{64\pi^{4} M^{3}}\,
\int_{0}^{\infty }\,dk\,k^{2}\,
\int_{k}^{\infty }\,d\omega \,\frac{1}{e^{\omega /T} - 1}
\nonumber 
\\ 
&=& \frac{1}{2880}\,\frac{\mu ^{2}T^{4}}{M^{3}} \,.
\end{eqnarray}
This calculation however ignores complicated logarithmic factors in
$r(\omega )$.

We numerically computed all terms including the logarithmic
factor in $r(\omega )$ along with $O[m^{2}]$ corrections.
It turns out that the total contribution is ten times larger 
than the analytic result above; in the $m\rightarrow 0$ limit,
\begin{equation}
n \approx  3.8\times 10^{-3}\,\frac{\mu ^{2}T^{4}}{M^{3}} \,.
\end{equation}
The main part of this large contribution comes from \\
$|\omega | < k$.
With a dimensionless constant introduced by $\mu = gM$, 
this gives, relative to the photon number density,
\begin{equation}
\frac{n}{T^{3}} \approx  4\times 10^{-13}\,
(\,\frac{g}{G_{F}m_{N}^{2}}\,)^{2}\,\frac{T}{M} \,.
\end{equation}
We wrote here the numerical value with $G_{F}$ the weak interaction constant of
mass dimensions $-\,2$
(\( \:
G_{F}m_{N}^{2} \approx 10^{-\,5} 
\: \)),
as if it were relevant to the neutron decay.

One may estimate the equal time temperature $T_{{\rm eq}}$ 
at which the power contribution becomes equal
to the Boltzmann suppressed number density, to give
\begin{equation}
\frac{T_{{\rm eq}}}{M} \approx \frac{1}{35} \,, \hspace{0.5cm} 
\frac{n}{T^{3}_{{\rm eq}}} \approx 1 \times 10^{-14} \,,
\end{equation}
taking as an example 
\( \:
\mu = 10^{-5}\,M \,,
\: \)
the weak interaction strength.
This number is in an interesting range to affect nucleosynthesis, but
we should keep in mind that we did not work out the relevant three
body decay, 
\( \:
n \rightarrow p + e + \bar{\nu }_{e} \,.
\: \)

The physical interpretation of the pole term is a conventional one
in terms of the remnant created by the inverse decay $\chi + \chi 
\rightarrow \varphi $, with all relevant particles on the mass shell,
hence suppressed kinematically by the Boltzmann factor $e^{-M/T}$.
On the other hand, the contribution from the continuous contour
can only be interpreted as remnant particles far off the mass shell
that may exist in thermal equilibrium.
The usual kinetic approach such as the Boltzmann-like equation is
based on the rates computed from S-matrix elements on the mass
shell and gives the Boltzmann suppressed abundance in equilibrium
for $M \gg T$. 
Our fully quantum mechanical approach yields
a different result.

We note that the local friction approximation is equivalent to
the pole model (with identification of
\( \:
\omega _{0}^{2} + \delta \omega ^{2} = (\Re z_{0})^{2} \,, \;
\eta = -\,2\Im z_{0}
\: \))
that ignores the continuum integral around the threshold, hence
the model fails to describe the off-shell remnant.

The effect of the off shell remnants
seems to play important roles at least in two places:
in nuclear matter and in the early universe.
We shall refer to \cite{nuclear medium review} for some recent
attempts to estimate medium effects in nucleus.

We shall mention another application of immediate interest in cosmology;
the heavy $X$ boson decay for GUT baryogenesis.
It has been argued \cite{baryogenesis review96} that there exists
a severe mass bound of order,
\begin{equation}
m_{X} > O[\alpha _{X}\,m_{{\rm pl}}] \approx 10^{16}\,{\rm GeV} \,, 
\end{equation}
to block the inverse process of the $X$ boson decay so that generation
of the baryon asymmetry proceeds with sufficient abundance of the parent
$X$ particles.
The usual estimate of the mass bound mentioned above however is based
on the on-shell Boltzmann equation.
More appropriate formula in this estimate is our remnant number density,
\begin{equation}
n_{X} \approx O[4\times 10^{-3}]\,g_{X}^{2}\,\frac{T^{4}}{m_{X}} \,.
\end{equation}
(In a more realistic estimate one should consider the $X$ boson decay
into quarks and leptons. But for an order of magnitude estimate difference
in statistics is not important.)
With the GUT coupling of $g_{X}^{2}/4\pi = 1/40$, the equal temperature
is roughly
\begin{equation}
T_{{\rm eq}} \approx \frac{M}{10} \,, \hspace*{0.5cm} 
\frac{n}{T^{3}_{{\rm eq}}} \approx 1 \times 10^{-4}
\,.
\end{equation}
Thus, at temperature of about a tenth of the $X$ mass the Boltzmann
suppressed formula is replaced by the power formula.
The kinematical condition for baryogenesis must be reconsidered in view of our
off-shell formula.

In summary, we extended previous works on the theory of quantum dissipation
in the linear harmonic environment, with emphasis on exact treatment
using the reduced density matrix.
When applied to the quantum system that decays via interaction with
the environment, 
the power law decay is a generic feature towards the asymptotic late
time limit.
The remnant fraction in the asymptotic limit has contribution from
particles off the mass shell that does not suffer from 
the Boltzmann suppression factor.


\begin{thebibliography}{99}
\bibitem{power comment}
It is known from any of reliable textbooks on quantum mechanics 
that the decay law of
unstable state can not be completely exponential, 
as for instance explained in detail in Goldberger and Watson, ref[13].
The nontrivial part is whether the power law persists with presence of
environments, not at zero, but at finite temperatures,
and if it does, how important it is from the point of a possible observation.
It is our view that 
the power law at any finite temperatures was not discussed in the past,
in generality that it really deserves.
\end{thebibliography}
\end{document}